# Selective Chemical Vapor Sensing with MoS$_2$ Thin-Film Transistors: Comparison with Graphene Devices

## R. Samnakay[1,2], C. Jiang[2], S. L. Rumyantsev[3,4], M.S. Shur[3] and A.A. Balandin[1,2,*]

[1]Phonon Optimized Engineered Materials (POEM) Center, Materials Science and Engineering Program, University of California – Riverside, Riverside, California 92521 USA

[2]Nano-Device Laboratory, Department of Electrical Engineering, Bourns College of Engineering, University of California – Riverside, Riverside, California 92521 USA

[3]Department of Electrical, Computer, and Systems Engineering, Center for Integrated Electronics, Rensselaer Polytechnic Institute, Troy, New York 12180, USA

[4]Ioffe Physical-Technical Institute, St. Petersburg 194021, Russia

**Abstract**

We demonstrated selective gas sensing with MoS$_2$ thin-film transistors using the change in the channel conductance, characteristic transient time and low-frequency current fluctuations as the sensing parameters. The back-gated MoS$_2$ thin-film field-effect transistors were fabricated on Si/SiO$_2$ substrates and intentionally aged for a month to verify reliability and achieve better current stability. The same devices with the channel covered by 10 nm of Al$_2$O$_3$ were used as reference samples. The exposure to ethanol, acetonitrile, toluene, chloroform, and methanol vapors results in drastic changes in the source-drain current. The current can increase or decrease by more than two-orders of magnitude depending on the analyte. The reference devices with coated channel did not show any response. It was established that transient time of the current change and the normalized spectral density of the low-frequency current fluctuations can be used as additional sensing parameters for selective gas detection with thin-film MoS$_2$ transistors.

R. Samnakay, C. Jiang, S. L. Rumyantsev, M.S. Shur and A.A. Balandin, UC-Riverside, RPI, Ioffe Institute (2014)

Two-dimensional (2D) layered materials have attracted significant attention owing to their unusual electronic and optical properties [1-4]. Among these material systems, a semiconducting $MoS_2$ is one of the most promising [5-6]. Each layer of $MoS_2$ consists of one sub-layer of molybdenum sandwiched between two other sub-layers of sulfur in a trigonal prismatic arrangement [7]. A direct band gap of a single-layer $MoS_2$ is ~1.9 eV [8-9]. Single-layer and few-layer $MoS_2$ devices have been proposed for electronic, optoelectronic and energy applications [1-4, 10]. Recently, $MoS_2$ film-based field-effect transistors were tested for sensing NO and $NO_2$, other gases, and water vapor [11-16]. The sensing signal utilized in these experiments was the relative change in the resistance, $\Delta R/R$. The devices with 2D channels are natural candidates for sensor applications due to the ultimately high surface-to-volume ratio and widely tunable Fermi-level position.

In this letter we report on selective detection of ethanol, acetonitrile, toluene, chloroform, and methanol vapors with the $MoS_2$ thin-film field-effect transistors (TF-FETs). The devices were intentionally aged for over a month period to verify their suitability for practical applications. In addition to the relative change in the source-drain current, $\Delta I_D/I_D$, we used the normalized spectral density of the low-frequency current fluctuations, $S_I/I_D^2$, and the characteristic transient time of the current as the sensing parameters (here $I_D$ is the source-drain current). Our results show that the aged $MoS_2$ devices perform better as the sensors in terms of their current stability, sensitivity to the analyte and reduced contributions of metal contacts to the noise level. Comparison with the graphene FETs reveals significant differences in the effects of exposure to chemical vapors on $\Delta R/R$ and $S_I/I_D^2$, suggesting differences in the physical mechanisms of low-frequency current fluctuations [17-19].

Thin films of $MoS_2$ were mechanically exfoliated from bulk crystals and transferred onto Si/$SiO_2$ substrates following the standard approach [1]. The thickness H of the films ranged from bi-layer to a few layers. Micro-Raman spectroscopy (Renishaw InVia) confirmed the crystallinity and thickness of the $MoS_2$ flakes after exfoliation. The spectroscopy was performed in the backscattering configuration under λ=488-nm laser excitation using an optical microscope (Leica) with a 50× objective. The excitation laser power was limited to less than 0.5 mW to avoid local heating. Figure 1 shows the schematic of $MoS_2$ TF-FET (a) optical microscopy image of a representative device (b) and Raman spectrum of the channel material (c). The observed Raman features at ~382.9 cm$^{-1}$ (E$^1_{2g}$) and ~406.0 cm$^{-1}$ (A$_{1g}$) are consistent with literature reports [20]. Analysis of the Raman spectrum indicates that this sample is 2-3 layer MoS$_2$ film. The thickness identification is based on the frequency difference, $\Delta\omega$, between the E$^1_{2g}$ and the A$_{1g}$ peaks. The increase in the number of layers in MoS$_2$ films is accompanied by the red shift of the E$^1_{2g}$ and blue shift of the A$_{1g}$ peaks [20].

[Figure 1: Schematic, microscopy and Raman]

Devices with MoS$_2$ thin-film channels were fabricated using the electron beam lithography (LEO SUPRA 55) for patterning of the source and drain electrodes and the electron-beam evaporation (Temescal BJD-1800) for metal deposition. The Si/SiO$_2$ (300-nm) substrates were spin-coated (Headway SCE) and baked consecutively with two positive resists: first, methyl methacrylate (MMA) and then, polymethyl methacrylate (PMMA). The resulting TF-FETs consisted of MoS$_2$ thin-film channels with Ti/Au (10-nm / 100-nm) contacts. The heavily doped Si/SiO$_2$ wafer served as a back gate. The majority of the bi-layer and tri-layer thickness MoS$_2$ devices had a channel length, L, in the range from 1.3 $\mu$m to 3.5 $\mu$m, and the channel width, W, in the range from 1 $\mu$m to 6 $\mu$m. Some of the devices were covered with 10-nm Al$_2$O$_3$ layer to serve as the reference samples in control experiments.

It is known that that defective and doped graphene has the greater sensitivity for CO, NO, NO$_2$, and other gases [21]. The same can possibly be true for thin film MoS$_2$ devices. In particular, it has been shown that the high density of edge states enhances the sensitivity. Therefore aged, i.e. more defective devices can be more attractive for gas sensing applications. In the present work we found that aged MoS$_2$ TF-FETs were more stable and had negligible contact contribution to the drain-to-source resistance. The aged MoS$_2$ devices were characterized by the on-to-off ratio of ~10$^4$, electron mobility $\mu$~0.5 cm$^2$/Vs and negligible contact resistance. The new, as fabricated, devices had mobility values in the range from 1 to 8 cm$^2$/Vs, which is typical for the back-gated MoS$_2$ TF-FETs [10, 22-23]. An estimate for the contact resistances was obtained by plotting the drain-to-source resistance, R$_{DS}$, vs. 1/(V$_G$-V$_{TH}$), and extrapolating this dependence to zero. Figure 2 shows a representative transfer current-voltage (I-V) characteristic for the aged MoS$_2$

TF-FET used in this study. For comparison, typical I-Vs for a graphene device are also shown. Analyzing I-Vs for $MoS_2$ thin films and graphene one can see possible implications for sensor operation: graphene device have much higher currents owing to graphene superb mobility while $MoS_2$ devices have better gating and on-off ration owing to $MoS_2$ band gap.

[Figure 2: Current – voltage characteristics of $MoS_2$ and graphene devices]

For testing the sensor operation, the vapors were produced by bubbling dry air through the respective solvents and diluting the gas flow with the dry air. The resulting concentrations were ~0.5 $P/P_o$, where P is the vapor pressure and $P_o$ is the saturated vapor pressure. When the sample is exposed to the vapor, the vapor molecules, which attach the channel surface, create negative or positive charges at the $MoS_2$ surface (see Figure 1 (a)). The latter depletes or enhances the electron concentration in the channel depending on the vapor species. For testing $MoS_2$ TF-FETs we selected three polar solvents: acetonitrile ($CH_3CN$ – polar aprotic), ethanol ($C_2H_5OH$ – polar protic), methanol ($CH_3OH$ – polar protic); as well as two non-polar solvents: toluene ($C_6H_5$-$CH_3$) and chloroform ($CHCl_3$).

Figure 3 shows the drain current as a function of time in $MoS_2$ TF-FET exposed to ethanol, methanol and acetonitrile, respectively (from top to bottom). For all chemical vapors, the measurements were conducted at the small drain-source voltage $V_D$=0.1 V. The high gate bias of $V_G$=60 V is explained by presence 300-nm thick $SiO_2$ layer in the back gate. One can see from Figure 3 that in all cases of polar solvents, the drain current increased upon exposure to the vapor and decreased after the vapor exposure was turned off. The results were reproducible for several switching on and off over in a month old $MoS_2$ TF-FETs tested over a period of few days. The time constants, $\tau$, for $I_D$ increase and decrease were different for each examined analyte. Note that the current of the reference sample – the same device with the channel coated by $Al_2O_3$ – did not reveal any changes.

[Figure 3: I-Vs for polar solvents]

Figure 4 presents drain current as a function of time in $MoS_2$ TF-FET exposed to chloroform and toluene, respectively (from top to bottom). To better illustrate a large range of the current change, Figure 4 also shows the data for chloroform in a semi-logarithmic scale. The exposure to the vapors of non-polar solvents has an opposite effect on $I_D$. The drain current reduces by more than two-orders of magnitude, almost completely switching the device off. We found also that the response of the $MoS_2$ transistors to some vapors demonstrates a memory effect: current remains small or even continue to decrease after the vapor flow is switched off. The current (resistance) is completely restored after gate and voltage biases are set to zero. The $\tau$ values were different for each analyte. Although not shown, no changes in current were observed for the reference samples. It is important to note that the data presented in Figures 3 and 4 were for devices intentionally aged by a month. Initially, the aging study was performed to verify that $MoS_2$ TF-FETs maintain their characteristic. Any practical sensor applications require that the FETs are operational for at least a month. Interestingly, we observed that the characteristics of the aged devices even improved in terms of their current stability and sensitivity.

[Figure 4: I-Vs for non-polar solvents]

We have recently demonstrated the use of the low-frequency current fluctuations of $I_D$ in graphene devices as an additional sensing parameter [24-25]. Some gases induce bulges at characteristic frequencies in the spectral density of the low-frequency current fluctuations $S_I/I_D^2$ of graphene FETs or change its average value. In order to test the same approach for $MoS_2$ TF-FETs we measured the low-frequency noise spectral density in $MoS_2$ TF-FETs exposed to open air and chemical vapors. The details of our low-frequency measurements have been reported by us elsewhere [26]. Figure 5 shows $S_I/I_D^2$ of $MoS_2$ TF-FET for different vapors. Red lines show the noise spectra measured in open air with intervals of a few day. The change of the noise spectra was found only as a result of the exposure to the acetonitrile vapor (indicated by green lines). The inset in Figure 5 magnifies the low-frequency part of spectra indicating good noise spectra measurements reproducibility. One can see that $S_I/I_D^2 \propto 1/f$ (here f is the frequency) without traces of bulges for all spectra including those measured under acetonitrile vapor. This is in a drastic contrast to graphene devices. In this sense, $S_I/I_D^2$ is a better sensing parameter for graphene rather than for few-layer $MoS_2$ films. However, the properly calibrated average level of $S_I/I_D^2$ can still be used for $MoS_2$ TF-FETs in combination with $\tau$ and $\Delta I_D/I_D$. Table I summarizes the data for examined chemical vapors including their type, dielectric constant $\varepsilon$ and two sensing parameters. In addition to the relative current change, $\Delta I_D/I_D$, we also show the ratio of the drain current under gas exposure to the initial current before exposure, $I_0$. The $I_D/I_0$ metric is illustrative for the situation when the drain current decreases. For comparison, the $\Delta I_D/I_D$ data for graphene FETs is also shown. As seen, the presence of the band gap in $MoS_2$, allows one to achieve a larger change in the current (orders of magnitude in some cases) than in graphene devices. For some vapors, it is possible to completely switch off $MoS_2$ by the gas. The latter makes $MoS_2$ very attractive for the gas sensing applications.

**Table I:** Chemical Vapors and Sensing Parameters of $MoS_2$ TF-FET

| Vapor | Type | $\varepsilon$ | $\tau$ (s) | $I_D/I_0$ $MoS_2$ FET | $\Delta I_D/I_D$ (%) $MoS_2$ FET | $\Delta I_D/I_D$ (%) Graphene FET |
|---|---|---|---|---|---|---|
| ethanol | polar protic | 24.6 | ~35 | 3.778 | +300 | - 50* |
| methanol | polar protic | 33.0 | ~20 | 3.714 | +280 | - 40* |
| acetonitrile | polar aprotic | 37.5 | ~130 | 1.625 | +60 | - 35* |
| chloroform | non-polar | 4.81 | ~ 550 | 0.231 | - 75 | - 25* |
| toluene | non-polar | 2.38 | ~900 | 0.012 | - 98 | +15* |

*The data for graphene is after S. Rumyantsev, et al., Nano Letters, 12, 2294 (2012) [24].

Although the exact mechanism of vapor molecule interactions goes beyond the scope of this work and requires atomistic simulations, one can observe certain trends from data in Table I, and Figures 3 and 4. The characteristic time constants for current change under the exposure to polar molecules are much shorter than those for non-polar molecules. The vapors of non-polar solvents, characterized by small ε, induce current quenching in the thin $MoS_2$ channel. The vapors of polar solvents with much larger ε, on contrary increase the electrical current in $MoS_2$ channel likely via inducing additional charges. Indeed, considering that ε for a few-layer $MoS_2$ is around 4 [27], the polar molecules would create a much larger dielectric mismatch with $MoS_2$.

The absence of bulges in $MoS_2$ TF-FETs, which is in contract to graphene FETs, can be related to different mechanisms of low-frequency noise in these two material systems. We have previously shown that low-frequency noise in $MoS_2$ thin films is similar to that in conventional semiconductors and can be described well by McWhorter model, which assumes that the dominant noise contribution comes from the number of careers fluctuations [26]. Graphene, similar to metals, reveals noise response, which is does not comply with the McWhorter model [17-19, 28]. Although the low frequency noise mechanism in graphene is still under debates [19] it is clear now that it is different from that in $MoS_2$. The latter makes the response of noise to the gas exposure to be also different in these two materials.

In conclusion, we demonstrated selective gas sensing with $MoS_2$ TF-FETs using the change in the channel current, characteristic transient time and low-frequency noise spectral density as the sensing parameters. The exposure to ethanol, acetonitrile, toluene, chloroform, and methanol vapors results in drastic changes in the source-drain current. The current can increase or decrease by more than two-orders of magnitude depending on the analyte type. The reference devices with coated channel did not show any response. Unlike graphene devices and thin-film $MoS_2$ transistors do not show characteristic bulges in the low-frequency current fluctuation spectra. The differences in the low-frequency noise response are likely related to differences in the noise mechanisms and require further study. The obtained results are important for practical applications of $MoS_2$ thin films and other van der Waals materials.

### *Acknowledgements*

The work at UC Riverside was supported by the Semiconductor Research Corporation (SRC) and Defense Advanced Research Project Agency (DARPA) through STARnet Center for Function Accelerated nanoMaterial Engineering (FAME). SLR acknowledges partial support from the Russian Fund for Basic Research (RFBR). The work at RPI was supported by the National Science Foundation under the auspices of the EAGER program. Authors thank Dr. V. Tokranov for the help with sample fabrication.

# References

[1] A. K. Geim and I. V. Grigorieva, Nature **499**, 419 (2013).

[2] D. Teweldebrhan, V. Goyal, and A. A. Balandin, Nano Lett. **10**, 1209 (2010).

[3] M. Chhowalla, H. S. Shin, G. Eda, L.-J. Li, K. P. Loh, and H. Zhang, Nat. Chem. **5**, 263 (2013).

[4] Z. Yan, C. Jiang, T. R. Pope, C. F. Tsang, J. L. Stickney, P. Goli, J. Renteria, T. T. Salguero, and A. A. Balandin, J. Appl. Phys. **114**, 204301 (2013).

[5] J. Heising and M. G. Kanatzidis, J. Am. Chem.Soc. **121**, 11720 (1999).

[6] Y. Kim, J.-L. Huang, and C. M. Lieber, Appl. Phys. Lett. **59**, 3404 (1991).

[7] J. L. Verble and T. J. Wieting, Phys. Rev. Lett. **25**, 362 (1970).

[8] S. W. Han, H. Kwon, S. K. Kim, S. Ryu, W. S. Yun, D. H. Kim, J. H. Hwang, J.-S. Kang, J. Baik, H. J. Shin, and S. C. Hong, Phys. Rev. B **84**, 045409 (2011).

[9] K. F. Mak, C. Lee, J. Hone, J. Shan, and T. F. Heinz, Phys. Rev. Lett. **105**, 136805 (2010).

[10] G. Eda, H. Yamaguchi, D. Voiry, T. Fujita, M. Chen, and M. Chhowalla, Nano Lett. **11**, 5111 (2011).

[11] H. Li, Z. Yin, Q. He, H. Li, X. Huang, G. Lu, D. W. H. Fam, A. I. Y. Tok, Q. Zhang, and H. Zhang, Small **8**, 63 (2012)

[12] D. J. Late, B. Liu, H. S. S. R. Matte, V.P. David and C. N. R. Rao, ACS Nano **6**, 5635 (2012).

[13] Q. He, Z. Zeng, Z. Yin, H. Li, S. Wu, X. Huang and H. Zhang, Small, **8**, 2994 (2012).

[14] F. K. Perkins, A. L. Friedman, E. Cobas, P. M. Campbell, G. G. Jernigan and B. T. Jonker, Nano Lett., **13**, 668 (2013)

[15] M. Shur, S. Rumyantsev, C. Jiang, R. Samnakay, J. Renteria, and A.A. Balandin, "Selective gas sensing with MoS$_2$ thin film transistors", IEEE Sensors 2014, Valencia, Spain, November 2-5, 2014.

[16] S.L. Zhang, H. H. Choi, H.Y. Yue and W. C. Yang, Curr. Appl. Phys., **14**, 264 (2014)

[17 M. Z. Hossain, S. Rumyantsev, M. S. Shur, and A. A. Balandin, Appl. Phys. Lett. **102**, 153512 (2013).

[18] S. L. Rumyantsev, D. Coquillat, R. Ribeiro, M. Goiran, W. Knap, M. S. Shur, A. A. Balandin, and M. E. Levinshtein, Appl. Phys. Lett. **103**, 173114 (2013).

[19] A. A. Balandin, Nat. Nanotechnol. **8**, 549 (2013).

[20] C. Lee, H. Yan, L. E. Brus, T. F. Heinz, J. Hone, and S. Ryu, ACS Nano **4**(5), 2695 (2010).

[21] K. R. Ratinac, W. Yang , S. P. Ringer and F. Braet, Environ. Sci. Technol., **44**, 1167 (2010).

[22 H. Wang, L. Yu, Y.-H. Lee, Y. Shi, A. Hsu, M. Chin, L.-J. Li, M. Dubey, J. Kong, and T. Palacios, Nano Lett. **12**, 4674 (2012).

[23] M. Xu, T. Liang, M. Shi, and H. Chen, Chem. Rev. **113**, 3766 (2013).

[24] S. Rumyantsev, G. Liu, M. S. Shur, R. A. Potyrailo and A. A. Balandin, Nano Lett., **12**, 2294 (2012).

[25] S. Rumyantsev, G. Liu, R.A. Potyrailo, A.A. Balandin and M.S. Shur, IEEE Sens. J. , **13**, 2818 (2013).

[26] J. Renteria, R. Samnakay, S. L. Rumyantsev, C. Jiang, P. Goli, M. S. Shur and A. A. Balandin, Appl.Phys. Lett. **104**, 153104 (2014).

[27] E. J. G. Santos and E. Kaxiras, ACS Nano, **7** (12), 10741 (2013).

[28] S. Rumyantsev, G. Liu, M. Shur, and A. A. Balandin, J. Phys.: Condens. Matter 22, 395302 (2010).

## FIGURE CAPTIONS

**Figure 1:** Schematic of the $MoS_2$ thin-film sensor with the deposited molecules that create additional charge (upper panel). Optical microscopy image of a representative $MoS_2$ TF-FET (middle panel). Raman spectrum of the $MoS_2$ thin-film channel showing the $E^1_{2g}$ and the $A_{1g}$ peaks. The increase in the number of layers in $MoS_2$ films is accompanied by the red shift of the $E^1_{2g}$ and blue shift of the $A_{1g}$ peaks. The energy difference, $\Delta\omega$, between $E^1_{2g}$ and the $A_{1g}$ peaks indicates that the given sample is a tri-layer $MoS_2$ film.

**Figure 2:** Current-voltage characteristics of the aged $MoS_2$ TF-FET used in the study. A typical graphene FET transfer characteristic is also shown for comparison in the same scale. Graphene reveals much higher current owing to superior electron mobility. $MoS_2$ TF-FET is characterized by better on-off ratio owing to its energy band gap.

**Figure 3:** Drain-source current versus time in $MoS_2$ TF-FET exposed to the vapors of polar solvents: ethanol (upper panel), methanol (middle panel), and acetonitrile (lower panel). The data were taken at the same gate voltage $V_G$=60 V and drain voltage $V_D$=0.1 V for each case. The reference sample – the same device with the $Al_2O_3$ coated channel – has not revealed any variations in current as shown in the upper panel

**Figure 4:** Drain-source current versus time $MoS_2$ TF-FET exposed to the vapors of non-polar solvents: chloroform (upper and middle panels) and toluene (lower panel). The data were taken at the same gate voltage $V_G$=60 V and drain voltage $V_D$=0.1 V for each case.

**Figure 5:** The normalized spectral density of the low-frequency source-drain current fluctuations measured for $MoS_2$ TF-FET in open air and under exposure to the vapors. The data were taken at the same gate voltage $V_G$=60 V and drain voltage $V_D$=0.1 V for each case. The spectral density of $MoS_2$ TF-FET under vapor exposure reveals 1/f dependence without any traces of bulges unlike graphene FETs. The inset magnifies the low-frequency part of the spectra.

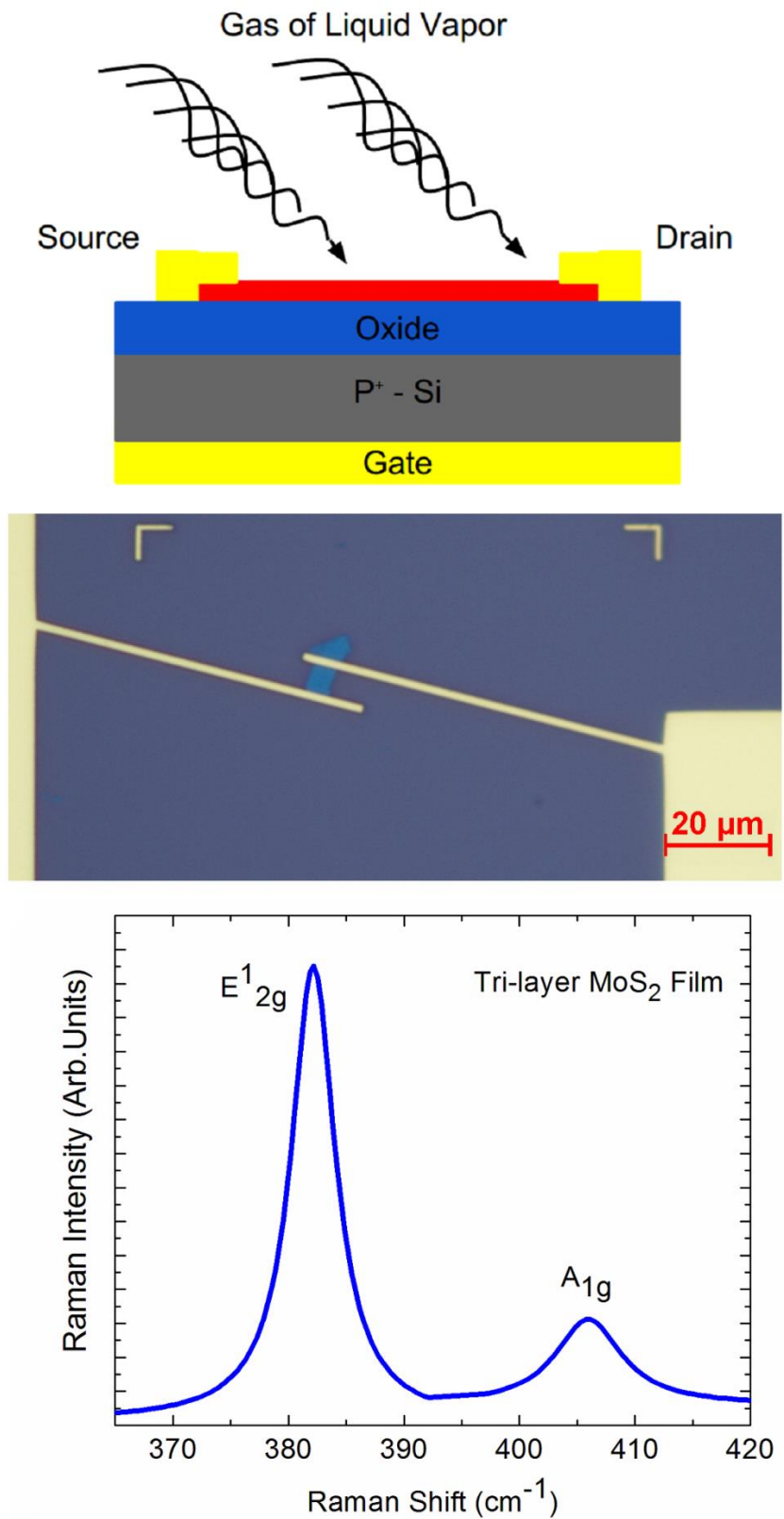

Figure 1

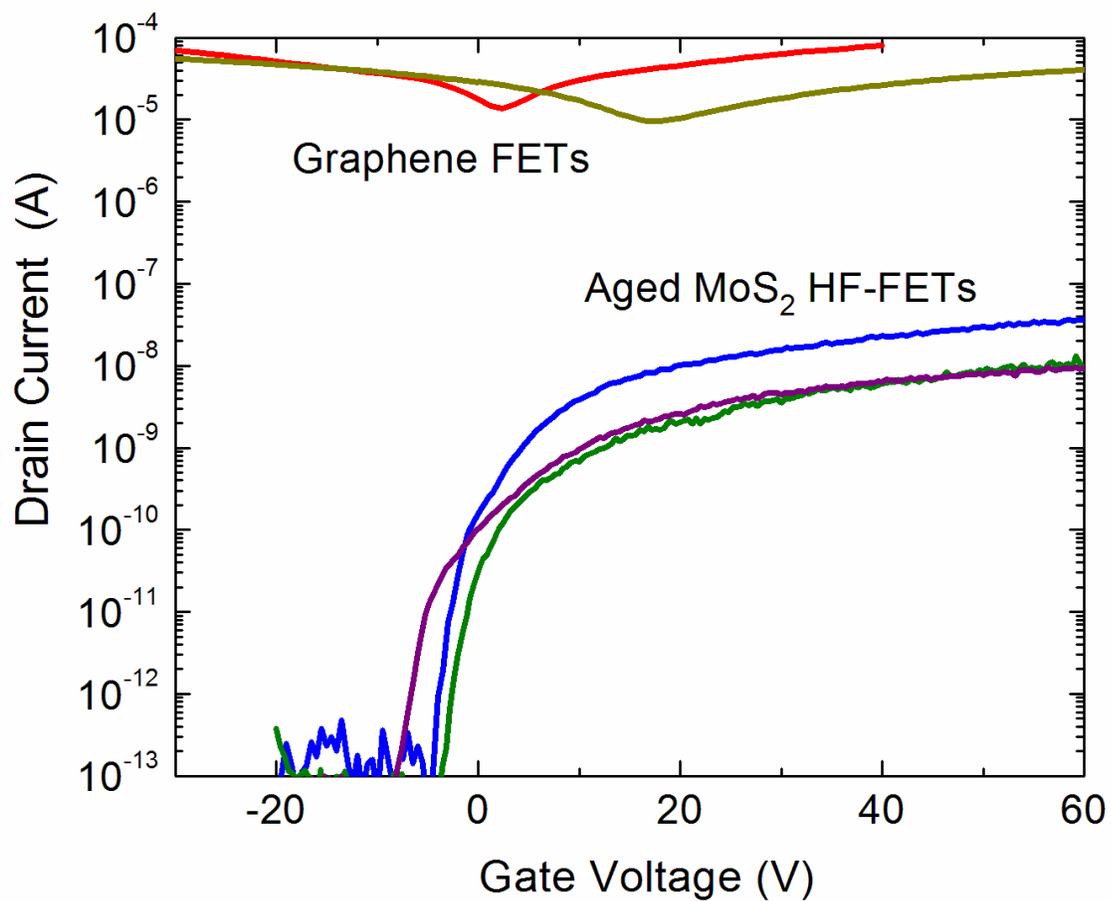

Figure 2

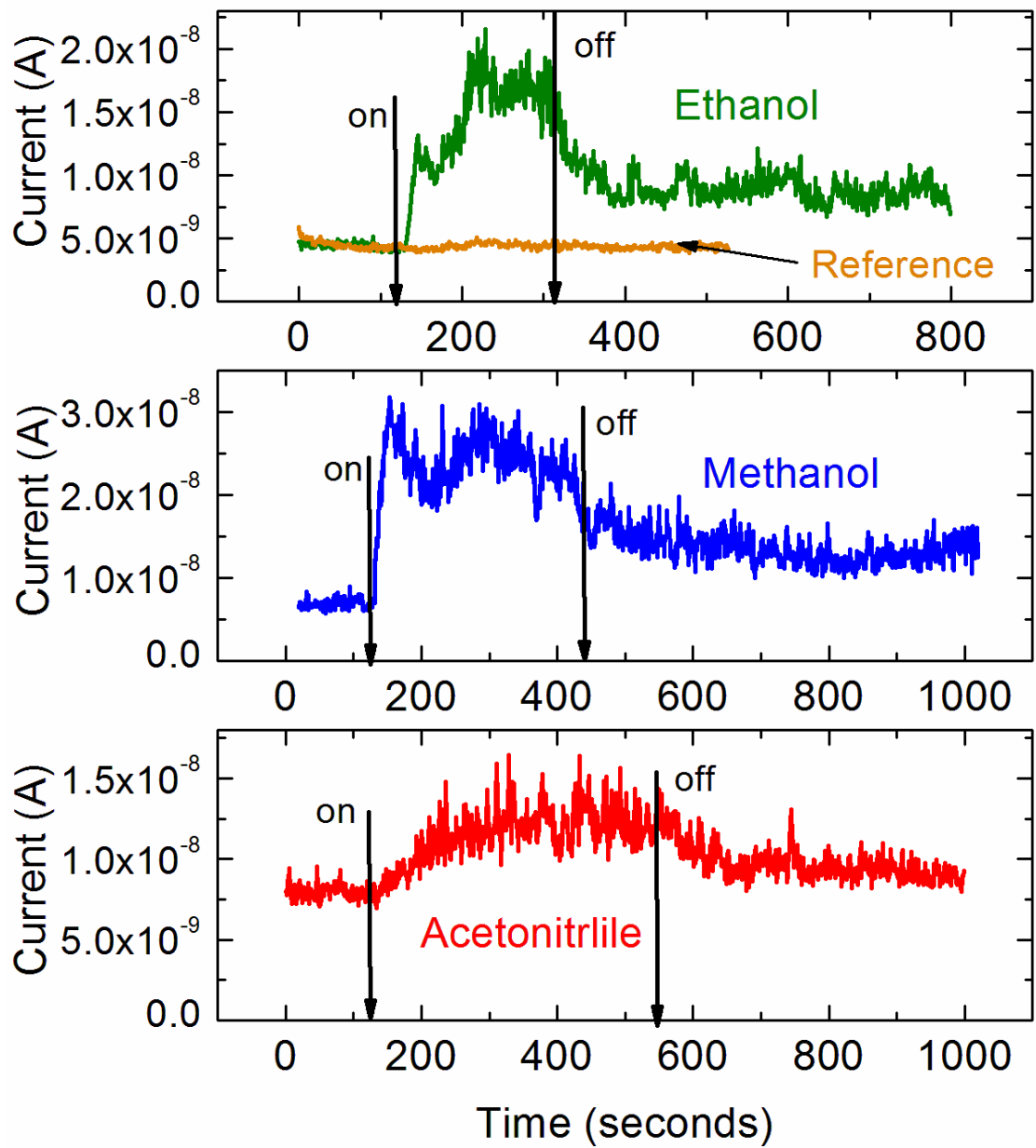

Figure 3

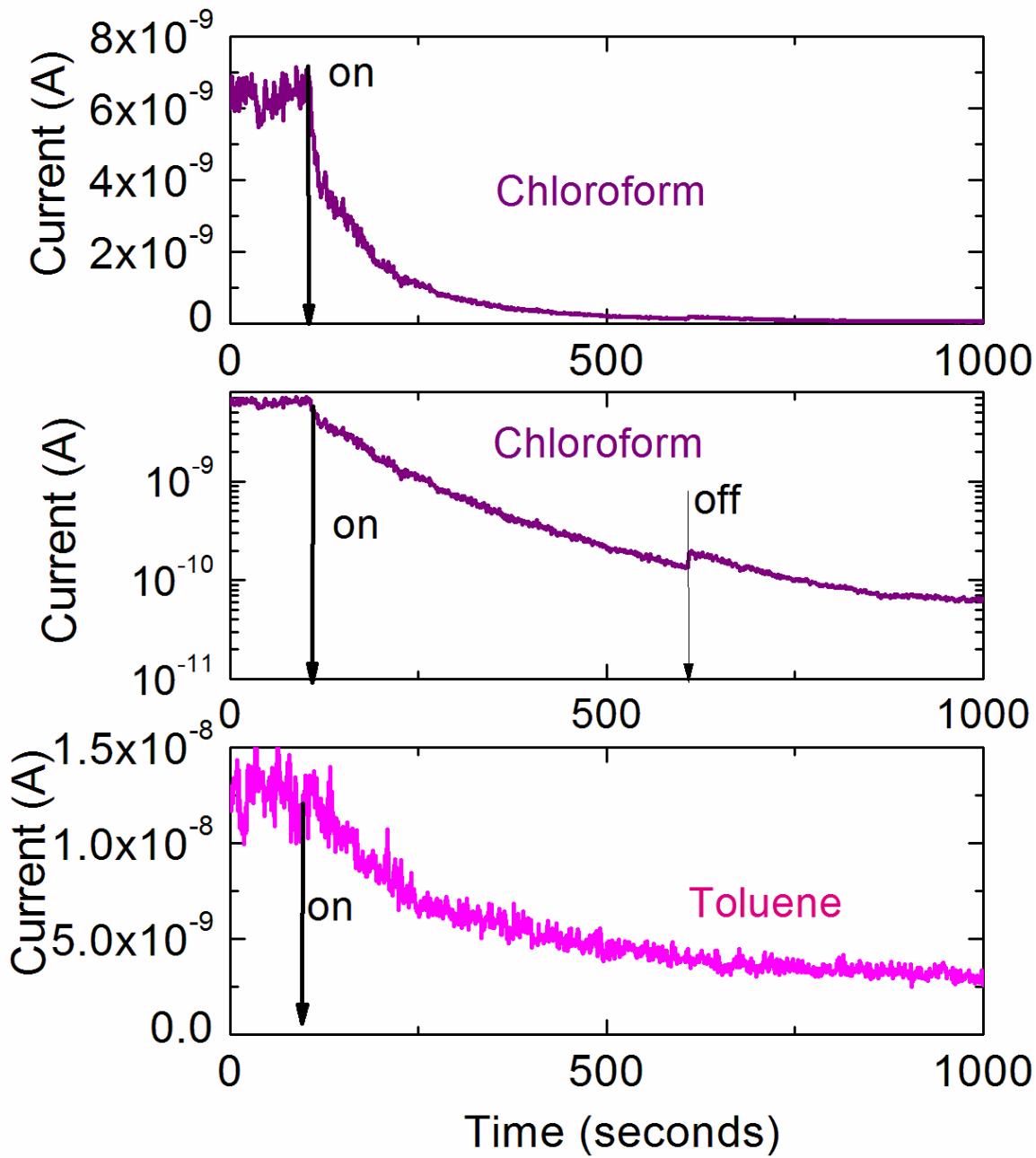

Figure 4

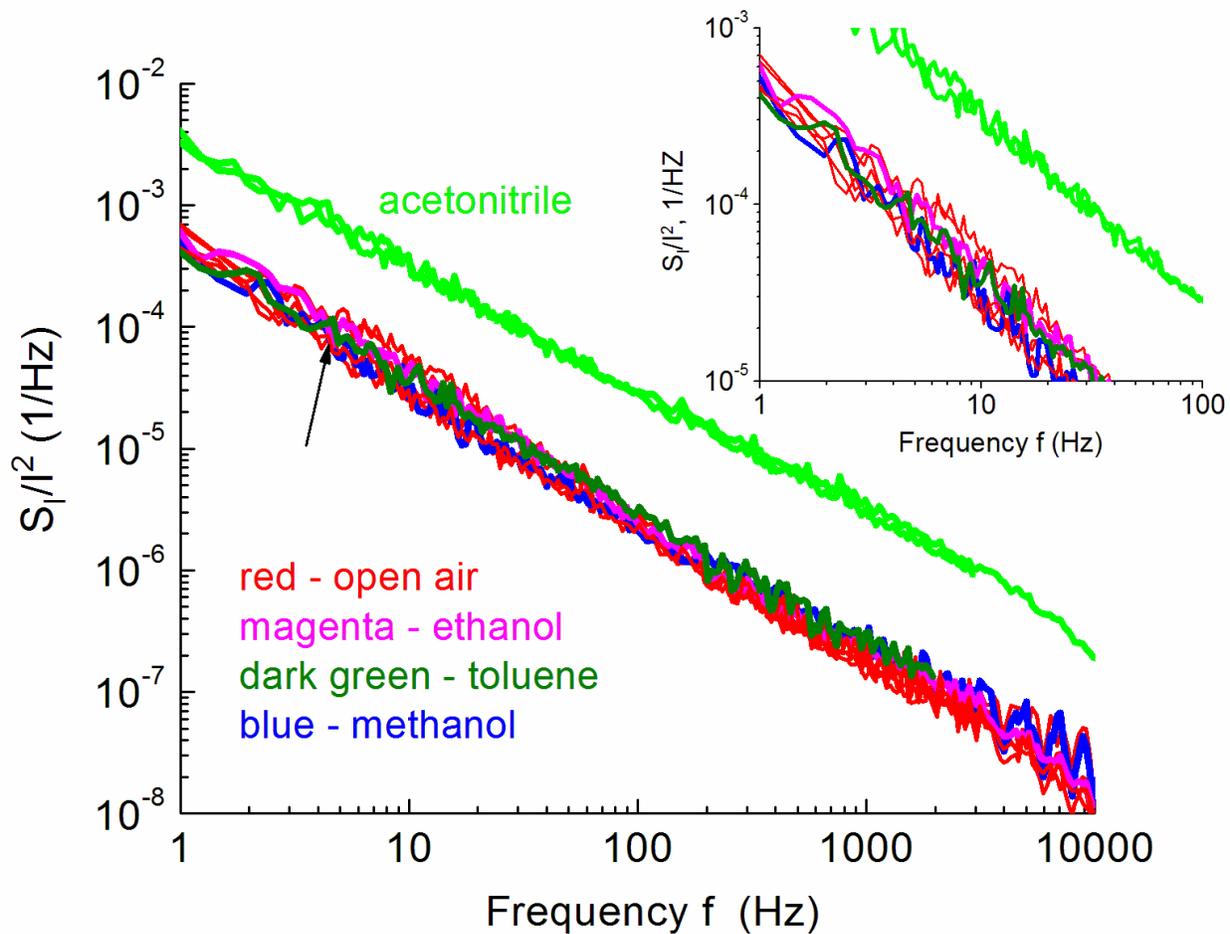

Figure 5